\documentclass[11pt,a4paper]{article}
\pdfoutput=1
\usepackage{jcappub}
\usepackage{color}
\usepackage[T1]{fontenc}
\usepackage{ifthen}
\usepackage{booktabs}
\newcommand{\beqra}{\begin{eqnarray}}
\newcommand{\eeqra}{\end{eqnarray}}
\newcommand{\beq}{\begin{equation}}
\newcommand{\eeq}{\end{equation}}

\title{On the proper kinetic quadrupole CMB removal and the quadrupole anomalies}
\author[a,b]{Alessio Notari}
\author[c]{and Miguel Quartin}
\affiliation[a]{Departament de F\'isica Fondamental i Institut de Ci\'encies del Cosmos,
Universitat de Barcelona, Mart\'i i Franqu\'es 1, 08028 Barcelona, Spain
}
\affiliation[b] {Dip. di Fisica, Universit\`a di Ferrara and INFN Sez. di Ferrara, Via Saragat 1, I-44100 Ferrara, Italy}
\affiliation[c]{Instituto de F\'\i sica, Universidade Federal do Rio de Janeiro, 21941-972, Rio de Janeiro, RJ, Brazil}

\abstract{It has been pointed out recently that the quadrupole-octupole alignment in the CMB data is significantly affected by the so-called kinetic Doppler quadrupole (DQ), which is the temperature quadrupole induced by our proper motion. Assuming our velocity is the dominant contribution to the CMB dipole we have $v/c=\beta=(1.231 \pm 0.003) \times 10^{-3}$, which leads to a non-negligible DQ of ${\cal O}(\beta^2)$. Here we stress that one should properly take into account that CMB data are usually not presented in true thermodynamic temperature, which induces a frequency dependent boost correction. The DQ  must therefore  be multiplied by a frequency-averaged factor, which we explicitly compute for several Planck CMB maps finding that it varies between $1.67$ and $2.47$. This is often neglected in the literature and turns out to cause a small but non-negligible  difference in the significance levels of some quadrupole-related statistics. For instance the alignment significance in the SMICA 2013 map goes from $2.3\sigma$ to $3.3\sigma$ with the frequency dependent DQ, instead of $2.9\sigma$ ignoring the frequency dependence in the DQ. Moreover as a result of a proper DQ removal, the agreement across different map-making techniques is improved.}


\begin{document}
\maketitle

\section{Introduction}

Since the observation of the anisotropies of the Cosmic Microwave Background  (CMB) by COBE and WMAP satellites several authors have pointed out the existence of possible anomalies with respect to the standard paradigm of an isotropic $\Lambda$CDM cosmological model with Gaussian adiabatic and nearly scale-invariant primordial perturbations (see~\cite{Bennett:2010jb,Ade:2013nlj} for an overview).  The first task when investigating possible anomalies to any well-established model is to ensure that all systematics are taken into account.

Among the many systematics involving a construction of a CMB map we focus here on those affected by our peculiar velocity with respect to the CMB, which induces both a Doppler and an aberration effect. The former's main effect is to introduce a large dipole. But the effects of a boost to a CMB map extend well beyond the dipole: aberration changes the arrival directions of photons and Doppler introduces a multiplicative factor and as a consequence both effects correlate the spherical harmonic coefficients in a non-trivial way (see for instance~\cite{Challinor:2002zh,Burles:2006xf,Notari:2011sb}).
It was in fact realized in~\cite{Kosowsky:2010jm,Amendola:2010ty,Notari:2011sb} that Planck would be able to detect this signature and obtain an independent measurement of our velocity, which was subsequently done in~\cite{Aghanim:2013suk}. The velocity found this way is in agreement with the one responsible for the dipole, reinforcing its interpretation as having a non-primordial origin. Moreover, our velocity influences at least 3 of the discussed CMB anomalies: the low-quadrupole, the hemispherical asymmetry and the quadrupole-octupole alignment, all of which we now briefly discuss.

The first one is part of the observed lack of power in the largest scales of the sky. It has been noted since COBE that assuming a power law spectrum the first multipoles of the CMB are rather low~\cite{Bennett:1996ce}. This remained true with WMAP~\cite{Hinshaw:2003ex,Spergel:2003cb,Tegmark:2003ve, Efstathiou:2003wr,Efstathiou:2003tv} and Planck~\cite{Ade:2013kta}, and the spectrum for $\ell \le 40$ is currently in tension with the higher multipoles at around 2.5--3$\sigma$. The amplitude and spectral index are in fact basically fixed by fitting the high multipoles while the low-$\ell$ are consistently below the average value. Among these large scales one of the most debated one is the quadrupole~\cite{Hinshaw:1996uq,Bond:1998zw,Spergel:2003cb} whose amplitude is $\simeq 200 (\mu K)^2$, almost 6 times lower than the theoretical expectation for Planck's $\Lambda$CDM parameters, which is around $1170 (\mu K)^2$~\cite{Ade:2013zuv}. Note however that the Doppler effect, in addition to a large contribution to the dipole, also produces a small kinetic contribution to the quadrupole, which has to be taken into account properly. Moreover, it was realized in~\cite{Kamionkowski:2002nd} and later also in~\cite{Chluba:2004cn,Sunyaev:2013coa}  that such a quadrupole contribution has a frequency dependence, which makes its subtraction more subtle. We will address this issue quantitatively later on in this short note.

The hemispherical asymmetry is related to an apparent modulation of the power of the CMB anisotropies in the sky. In particular, by splitting the CMB in two hemispheres there seems to be in WMAP~\cite{Eriksen:2003db,Hansen:2004mj,Hansen:2004vq,Eriksen:2007pc,Hoftuft:2009rq} and also in Planck~\cite{Ade:2013nlj} an indication for a power asymmetry  along the direction $(l,b) = (225^\circ,1^\circ)$. In v1 of~\cite{Ade:2013nlj} such asymmetry was claimed to extend with a large amplitude up to $\ell=1500$; at the same time, it was also noted in~\cite{Aghanim:2013suk} that the Doppler effect  due to our motion should induce a $0.25\%$ power asymmetry. Later we have shown quantitatively in~\cite{Notari:2013iva} that \emph{both} Doppler \emph{and} aberration induce a power asymmetry and that it grows at small scales, reaching $0.6\%$ at high $\ell$ and that the subtraction of such effect is crucial in the estimation of the  asymmetry if considering $\ell$ up to about a thousand. In fact in revisions of~\cite{Ade:2013nlj} and in~\cite{Quartin:2014yaa} it was shown that the significance at high $\ell$ disappears after such subtraction so that a scale-independent power asymmetry is now actually excluded by Planck and a dipolar modulation of power can survive at most up to $\ell\lesssim 600$, though with a marginal significance of less than $3 \sigma$ (which also agrees with earlier results in~\cite{Hansen:2008ym} for WMAP).

Finally, it seems that the three lowest multipoles might possibly not be randomly distributed in the sky as there seems to be some alignment between them in the WMAP data~\cite{Tegmark:2003ve, Bielewicz:2004en}. In fact a direction for the quadrupole and octupole can be defined either by maximizing the analogous quantity to the angular momentum~\cite{deOliveiraCosta:2003pu} or by using the so-called multipole vectors~\cite{Copi:2003kt}. In both cases they are found to be extremely close to each other, which is unlikely in a  Gaussian isotropic distribution: the significance of such an alignment has been estimated to correspond to a $p$-value of about $0.1\%$ (see {\it e.g.}~\cite{Copi:2013jna}). With the release of the Planck data such an alignment seemed at face value to have turned to be less significant~\cite{Ade:2013nlj} to a $p$-value of $1\%-2\%$, but it has been pointed out~\cite{Copi:2013jna} that this has to be corrected for the Doppler effect, which induces a velocity dependent quadrupole (kinetic Doppler quadrupole, DQ) and this brings up again the significance of the alignment to a  level similar to the one found in WMAP of about $0.05\%-0.3\%$; note moreover that such common direction seems to be also correlated with the dipole direction, giving similar significance levels for a statistic which involves $\ell=1,2,3$ (see Table 5 and 7 of~\cite{Copi:2013jna}).

However, we stress here that due to the way in which the CMB  data are presented, both by WMAP and Planck, such a correction is actually frequency-dependent, as it has been already shown in~\cite{Kamionkowski:2002nd} and similarly to what has been discussed in~\cite{Aghanim:2013suk} in a different context. This results in an effective multiplicative correction factor for each frequency map. This factor is very different from 1 for the single frequencies, being especially large at high frequencies, and therefore it has a potentially large effect in a linearly combined temperature map, such as the Planck SMICA and NILC maps. To our knowledge such a frequency dependent factor for the DQ has only been quantitatively considered  in~\cite{Ade:2013nlj},  and it has been estimated to be of 1.7 for both the SMICA and NILC maps and then it has been applied to the Wiener filtered maps of the Planck collaboration  to show that the alignments have a significance of about $1\%$. Other groups~\cite{Copi:2013jna,Rassat:2013caa,Rassat:2014yna,Polastri:2015rda} have used instead just a factor of 1, which is significantly off (in one map the real value it is over 2.5, see below) and leads to a non-negligible modification of their results.  In this note we make an independent estimate of this factor for several maps, including two which are not constructed by the Planck collaboration.

\section{Frequency dependence of DQ in CMB Experiments}

Both WMAP and Planck collaborations are showing at each frequency intensity maps which are assumed to be proportional to the CMB temperature, through a linear series expansion.  However the exact relation between the intensity measured in our frame and the rest-frame CMB temperature contains a frequency dependent factor. Let us briefly recall how such a factor arises for the DQ, similarly to the treatment of~\cite{Kamionkowski:2002nd}.

A boost with a velocity $\,\mathbf{v}/c=\boldsymbol{\beta}\,$ has two effects on an image of the sky: Doppler effect and aberration. The first effect is a multiplicative direction dependent factor while the second effect changes the apparent arrival direction of photons~\cite{Peebles:1968}: %,Challinor:2002zh,Burles:2006xf
\begin{equation}
    T'(\boldsymbol{\hat{n}}')=
    \frac{T(\boldsymbol{\hat{n}})}{\gamma(1-\boldsymbol{\beta} \cdot \boldsymbol{\hat{n}}')} \,, \label{Dopplerab}
\end{equation}
where $\boldsymbol{\hat{n}'}$ is the (aberrated) arrival direction of the photons and $T'(\boldsymbol{\hat{n}'})$ is the temperature observed in our reference frame. Here we assume the CMB temperature in the rest frame $T(\boldsymbol{\hat{n}'})$ to be the sum of a homogenous component plus perturbations  in this way:
\begin{equation}
    T(\boldsymbol{\hat{n}})=T_0  + \varepsilon \, \delta T(\boldsymbol{\hat{n}}) \,,
\end{equation}
where by convention we assume $\varepsilon=10^{-5}$ and so  $ \delta T(\boldsymbol{\hat{n}})$ is of order 1.

If a map is completely homogeneous ($\varepsilon=0$) aberration has no impact,  and so its leading effect is of order $\varepsilon \, \beta$. However the Doppler effect is non-trivial even if a map is completely homogeneous and in fact it induces a dipole of order $\beta$ and an $n^{\rm th}$-pole of order $\beta^n$, as  can be seen by expanding the multiplicative factor in eq.~\eqref{Dopplerab}. Therefore, if our velocity is $\beta \approx 10^{-3}$, there is an induced dipole of order $10^{-3}$ and a kinetic Doppler quadrupole (DQ) correction of order  $10^{-6}$, which is not negligible. All other effects are at most of order of $\beta \varepsilon$ or $\beta^3$ and therefore negligible in the discussion of the alignments (but important for other purposes, such as measuring our velocity through non diagonal correlations between multipoles, as it has been shown in~\cite{Kosowsky:2010jm,Amendola:2010ty,Notari:2011sb, Aghanim:2013suk}).

As we already mentioned, WMAP and Planck frequency maps~\cite{Ade:2013sjv,Ade:2013hta,Adam:2015wua,Adam:2015tpy} are not produced in terms of temperature, but in terms of intensities multiplied by a proportionality factor and so the boost induces a frequency-dependent correction. An observer in our frame sees in fact an intensity at a given frequency $\nu'$ given by:
\begin{equation}
    I'(\nu') = I (\nu) \left( \frac{\nu'}{\nu}\right)^3 = \frac{2 \nu'^3}{e^{\frac{\nu }{T(\boldsymbol{\hat{n}})}}-1} \,.
\end{equation}
where $ \nu= \nu' \gamma(1-\boldsymbol{\beta} \cdot \boldsymbol{\hat{n}}') $. In the $\beta=0$ limit the two frames coincide and fluctuations in Intensity are given at first order in $\varepsilon$ by:
\begin{equation}
    \delta I(\nu)\,\approx\, \frac{2 \nu ^4   e^{\frac{\nu }{T_0}}}{T_0^2 \left(e^{\frac{\nu }{T_0}}-1\right)^2}  \varepsilon  \, \delta T(\boldsymbol{\hat{n}})
    \,\equiv\, K  \varepsilon \, \frac{\delta T(\boldsymbol{\hat{n}})}{T_0} \,.
    \label{K}
\end{equation}

The above approximate equation is used by the WMAP and Planck collaborations to \emph{define} temperature as $\delta I(\nu)/K$, which  differs from the real thermodynamic temperature. We will refer to $\delta I(\nu)/K$ here as \emph{linearized temperature}. We can in fact now write down the expansion at second order in $\beta$ and first order in $\varepsilon$ of the linearized temperature:
\begin{equation}\label{eq:lin-temp}
    \frac{\delta I'(\nu')}{K} \,=\, \varepsilon \, \frac{\delta T(\boldsymbol{\hat{n}})}{T_0} + \beta z   + \beta ^2 z^2 Q(\nu') -   \frac{1}{2} \beta ^2\,,
\end{equation}
where we have discarded terms of order $\beta\, \varepsilon$ or higher and where $z=\boldsymbol{\hat{\beta}}\cdot \boldsymbol{\hat{n}}$ and
\begin{equation}
    Q(\nu')=\frac{\nu'}{2 T_0}  \coth \left(\frac{\nu' }{2 T_0}\right) \,.
\end{equation}
in agreement with~\cite{Kamionkowski:2002nd}.\footnote{Note that there is a typo in the derivation of Eq. (2) of~\cite{Kamionkowski:2002nd}: $\nu = \gamma \nu'(1 + \beta \mu)$ should read $\nu = \gamma \nu'(1 + \beta \mu')$.} So, in addition to a dipole correction we also have a frequency dependent $z^2$ (quadrupole) correction and constant shift to the monopole. Using the fact that $T_0 = 56.8 \, {\rm GHz}$ we get that the correction factor $Q(\nu)$ has for Planck a value of about $1.5$ for the $143\,  {\rm GHz}$ channel and of about $2$ for the $217\,  {\rm GHz}$ channel, which are the main two channels which contain cosmological signal. However, in order to remove foregrounds all CMB maps come from combinations of more frequencies, so we perform here an explicit computation of the overall effective $Q_{\rm eff}$ for specific maps.

In this work we analyze 4 different CMB maps: 2 produced by the Planck Collaboration (SMICA 2013 and NILC 2013) and 2 produced by an independent group~\cite{Bobin:2014mja}. For both SMICA 2013 and NILC 2013  Planck provided inpainted maps where the inpainted regions are generated with a constrained Gaussian realization which assumes isotropy~\cite{BenoitLevy:2013bc}. Although it consists of only a small $3\%$ fraction of the sky, it is interesting to assess the effect of the inpainting assumptions, especially since the fiducial cosmology assumed in the inpainting is not discussed in detail in~\cite{Ade:2013hta}.  Using the public Planck inpainted maps~\cite{Ade:2013ktc} we find results that agree with the ones in~\cite{Copi:2013jna}, where a different inpainting technique was used.

It should be noted that the inpainting masks are smaller than the so-called confidence masks of Planck, which cover $11\%$ and $8\%$ of the sky for SMICA and NILC, respectively. So there could be some residual foreground contamination on these inpainted maps. Probably for this reason, Planck collaboration also analyzed an alternative SMICA map with a much larger $27\%$ mask, inpainted with a Wiener filter technique~\cite{Ade:2013nlj}. We could not compute the  significance of this map  because it is not publicly available. Instead, for SMICA and NILC we consider also the alternative maps provided in~\cite{Rassat:2014yna}, the only  two differences being that the masked region are inpainted with a different technique and and that for NILC the inpainted region correspond to the $8\%$ confidence mask instead of the $3\%$ one. We dub these maps SMICA (R14) and NILC (R14).

In what concerns the quadrupole and octupole, all 4 maps can be summarized as a simple linear combination of the different frequency channels with a set of weights, which sum to 1. In all cases except one the weight-vectors contain 9 entries and make use of all 9 Planck channels; the remaining map has a weight-vector containing 14 entries, due to inclusion also of WMAP's 5 channels. Once we have the weights $w_{\nu_i}$, we just compute:
\begin{equation}
    Q_{\rm eff} \,=\, \sum_i Q(\nu_i) w_{\nu_i}\,. \label{Qeff}
\end{equation}
We will quote the weights in order of increasing channel frequency, so that the 5 WMAP entries correspond to the frequencies $(23,30,40,60,90)$ GHz and the 9 Planck ones to the ($30,44,70,100,143,217,353,545,857$) GHz channels.

Given these weights, there is an important \emph{caveat}: the low frequency channels (the ones measured by the LFI instrument) of the 2013 Planck release have been already treated in such a way to remove the DQ, but a bug in the code removed only \emph{half} of the frequency-independent effect (see the Appendix of~\cite{2014A&A...571A...5P}). Therefore for the LFI channels we apply $Q(\nu) \rightarrow Q(\nu)-0.5 $. The high frequency channels (HFI) instead have not been treated at all, as described in Appendix of~\cite{Ade:2013eta}.

For WMAP alone the various channels are all at quite low frequency and so the $Q$ factors are between 1.01 and 1.22. In this case the subtraction with $Q=1$ obtained in~\cite{Copi:2013jna}  gives already a good approximation of the correct result.

For the Planck SMICA map, the weights provided in both Figures D.1 of~\cite{Ade:2013hta,Adam:2015tpy} are not the final numbers we are interested in, because they are in fact a product of 3 different functions: the actual CMB temperature weights, the conversion factors from Antenna Temperature to linearized temperature ({\it i.e.} the ratio between $K$ of eq.~\eqref{K} and its Rayleigh-Jeans small $\nu$ limit) and a beam and HEALPix\footnote{\url{http://healpix.sourceforge.net/}}~\cite{Gorski:2004by}  pixel window function correction~\cite{Ade:2013hta}. For the quadrupole and octupole, the last term is irrelevant. We obtained the linearized temperature weights by multiplying the unit conversion factors to the carefully inspected values of Figure D.1 of~\cite{Ade:2013hta}
\begin{equation}
    \mathbf{w}_{\rm SMICA13}\,=\,\{0.01,-0.094,-0.113,-0.256,1.184,0.436,-0.174,0.005,-8\times10^{-6}\}\,.
\end{equation}
As a consistency check such numbers now sum to 1 to good precision, as they should. Keeping in mind the different units, $\mathbf{w}_{\rm SMICA13}$ also agrees very well with the values read independently in a similar way by~\cite{Larson:2014roa}. We could do the same with SMICA 2015, but it is unclear how the DQ was treated there. In fact in~\cite{Adam:2015tpy}  it is stated that the DQ has already been subtracted and this is clearly visible in the difference of the 2013 and 2015 maps, but it is not clearly specified if the correct $Q(\nu)$ factors have been applied to the single frequency channels. Moreover, information regarding the calibration of the 2015 LFI channels has not yet been released.

For the Planck NILC map  only the first needlet band, which has a constant weight across the sky, is relevant. The weight vector for NILC 2015 map was obtained by careful inspection of Figure B.2 in~\cite{Adam:2015tpy} obtaining:
\begin{equation}
    \mathbf{w}_{\rm NILC15}\,=\,\{0.05, -0.055, -0.7, 0.25, 1.51, 0.03, -0.09, 0.005, 0\}\,.
\end{equation}
Unfortunately the weights for NILC 2013 are not reported in the Planck papers, so we cannot confirm the value of 1.7 obtained in~\cite{Ade:2013nlj}. In fact, the value obtained in 2015 differs from 1.7, which indicates that the weights changed between both releases. Thus, for the particular case of NILC we use this quoted value from Planck.

The LGMCA maps~\cite{Bobin:2014mja} uses a component separation method which is claimed to be able to reconstruct the full sky without the need of a mask nor inpainting. Specifically the authors have released two maps: the PR1 map, which uses the 9 Planck channels and the WPR1 which uses both the WMAP 5 channels and the Planck 9 channels.
The corresponding weights could be obtained from the map-construction code which is publicly available~\cite{cosmostat-webpage}:
\begin{equation}
    \mathbf{w}_{\rm PR1}\,=\,\{0.087, 0.087, -1.186, -0.668, 2.093, 0.885, -0.306, 0.008, -7\times10^{-6}\}\,,
\end{equation}
and
\begin{equation}
\begin{aligned}
    &\mathbf{w}_{\rm WPR1, WMAP} \,=\,\{0.062, -0.277, 0.227, 0.191, 0.413\}\,, \\
    &\mathbf{w}_{\rm WPR1, Planck} \,=\,\{0.147, -0.45, -0.827, 0.254, 1.412, -0.064, -0.09, 0.002, 6\times10^{-6}\}\,.
\end{aligned}
\end{equation}
Note, again, that such PR1 and  WPR1 weights respectively sum to unity. In such case we find that the factor is  2.47 and 2.05, respectively for PR1 and WPR1.

The final correction factors  at $\ell=2$  range from 1.67  to 2.47 and are given in Table~\ref{tab:alignments}, together with the change on the significance of the alignments, using several statistics. It is important to stress that individual $Q(\nu)$ for the higher frequencies have much larger values for Planck:  for instance they are  3.1 and 4.8 for the 353 GHz and 545 GHz Planck channels, and $Q_{\rm eff}$ is the result of cancellations among large contributions. Note also that we have checked that in the new release of Planck 2015 the weights have changed by a large amount, but as we already said we do not know if the correct overall $Q_{\rm eff}$ has been already applied or not to the maps; this will probably be clear once the LFI calibration paper is released.

Note that in principle $Q(\nu)$ should be averaged over the bandpasses for each channel with an appropriate weight. As a simple check we have performed a weighted average with a step function between some minimal and maximal frequency. We have found values for the minimal and maximal frequency for each HFI channel in Table 3 of~\cite{Ade:2013fta} and for LFI in~\cite{Zonca:2010fx} (although for LFI this is completely irrelevant). We find a  $1\%$ difference in the final $Q_{\rm eff}$, with respect to our previous procedure, so we have neglected such corrections in what follows.

\section{Summary of our Results}

Once we have the correct $Q_{\rm eff}$ factors we have then computed the significance of the alignments using three types of statistics, which make use of the multipole vectors (see~\cite{Copi:2013jna} and references therein). The multipole vectors are a different way of expressing the same information as the $a_{\ell m}$, which consists in one amplitude and $\ell$ unit vectors (which sums to $2\ell+1$ real quantities). From the multipole vectors one can build the area vectors ${\bf w}_j$, which are cross products of the multipole vectors.
One can also define instead directions $\boldsymbol{\hat{n}}_\ell$ of maximal ``angular momentum''
\begin{equation}\label{eq:ang-momentum}
    \sum_m m^2 \big|a_{\ell m}\big|^2\,,
\end{equation}
one for each $\ell$~\cite{deOliveiraCosta:2003pu}. For $\ell =2$ in fact $\boldsymbol{\hat{n}}_2 = {\bf w}_2/|{\bf w}_2|$, but for higher multipoles the relationship is more complicated; nevertheless they both probe similar effects~\cite{Copi:2005ff}. From the computational point of view, however, it is much simpler to compute the multipole vectors, as these can be found simply by computing the roots of a polynomial built using the $a_{\ell m}$'s of a given $\ell$~\cite{Helling:2006xh}.
For $\boldsymbol{\hat{n}}_\ell$, on the other hand, the standard practice is to rotate the $a_{\ell m}$'s in many directions (often millions) on the sky and compute  explicitly the ``angular momentum'' in each case. In order to compute $\boldsymbol{\hat{n}}_3$ we made use of the code developed in~\cite{Rassat:2014yna} and made publicly available~\cite{cosmostat-webpage}. We also cross-checked that $\boldsymbol{\hat{n}}_2 = {\bf w}_2/|{\bf w}_2|$ was found to hold comparing both techniques.

The first statistics is the simple relative angle between $\boldsymbol{\hat{n}}_2$ and $\boldsymbol{\hat{n}}_3$, where we assign a  probability assuming a flat distribution of $\boldsymbol{\hat{n}}_2 \cdot \boldsymbol{\hat{n}}_3$, which corresponds to random orientation of the two unit vectors.
The other two statistics that we used are the S and T variables~\cite{Copi:2013jna} which are defined instead with respect to a particular direction $\boldsymbol{\hat{n}}$ and they involve the area vectors of the quadrupole and of the octupole as follows:
\begin{equation}
    S=\frac{1}{n}\sum_{j=1}^n A_j\,, \qquad T=1-\frac{1}{n}\sum_{j=1}^n (1-A_j^2)\,, \
\end{equation}
where $A_j \equiv |{\bf} \mathbf{w}_j \cdot  \boldsymbol{\hat{n}}|$. In the present treatment we will only use as a specific direction $\boldsymbol{\hat{n}}$ the direction of the dipole, which has been shown to be the one with most significant alignment according to the analysis of~\cite{Copi:2013jna}. In other words we test the hypothesis that the dipole, quadrupole and octupole are all aligned.

\begin{figure}[t!]
\begin{minipage}[c]{\textwidth}
    %\centering
    \includegraphics[width=0.5\textwidth]{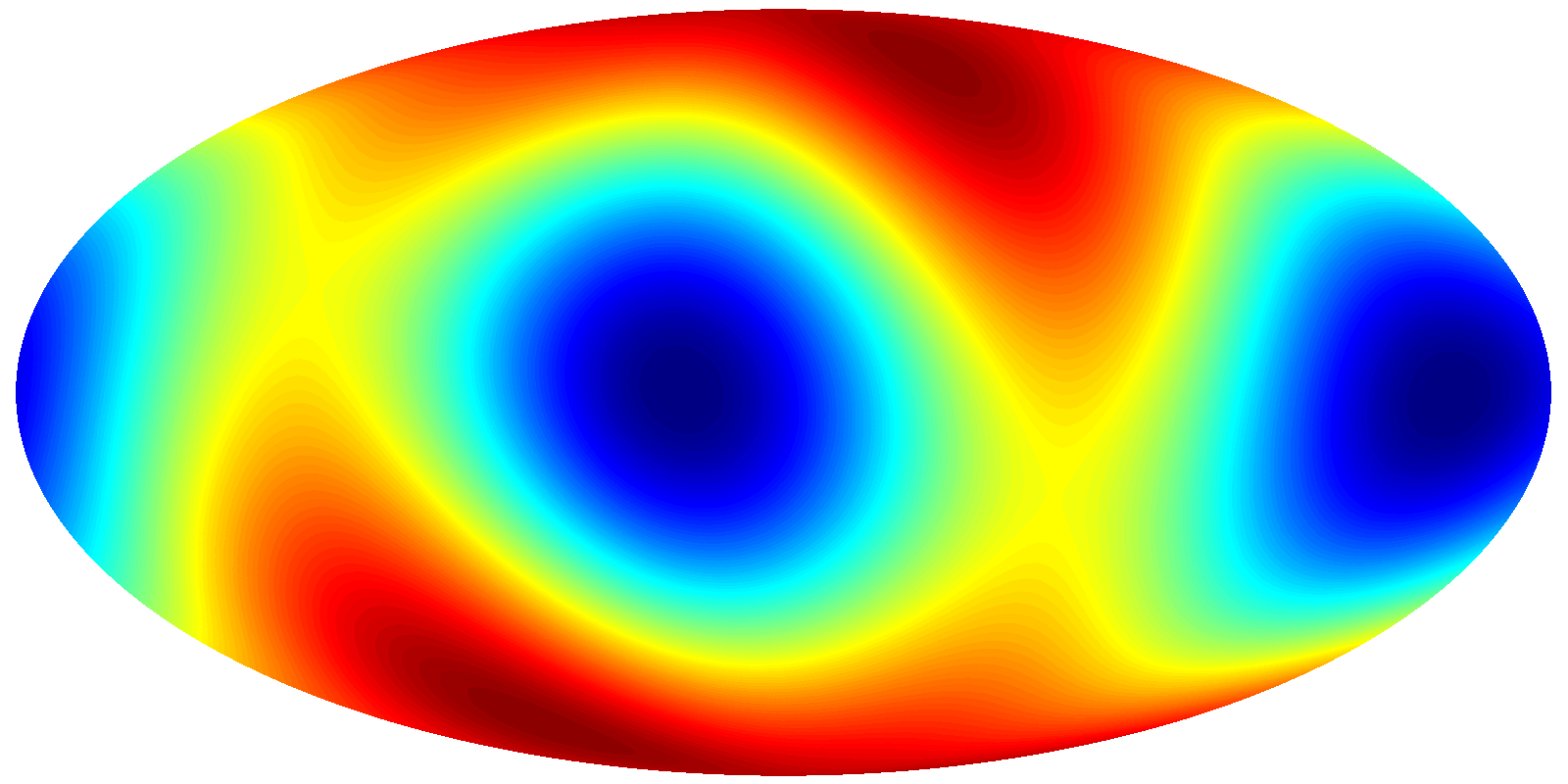}
    \includegraphics[width=0.5\textwidth]{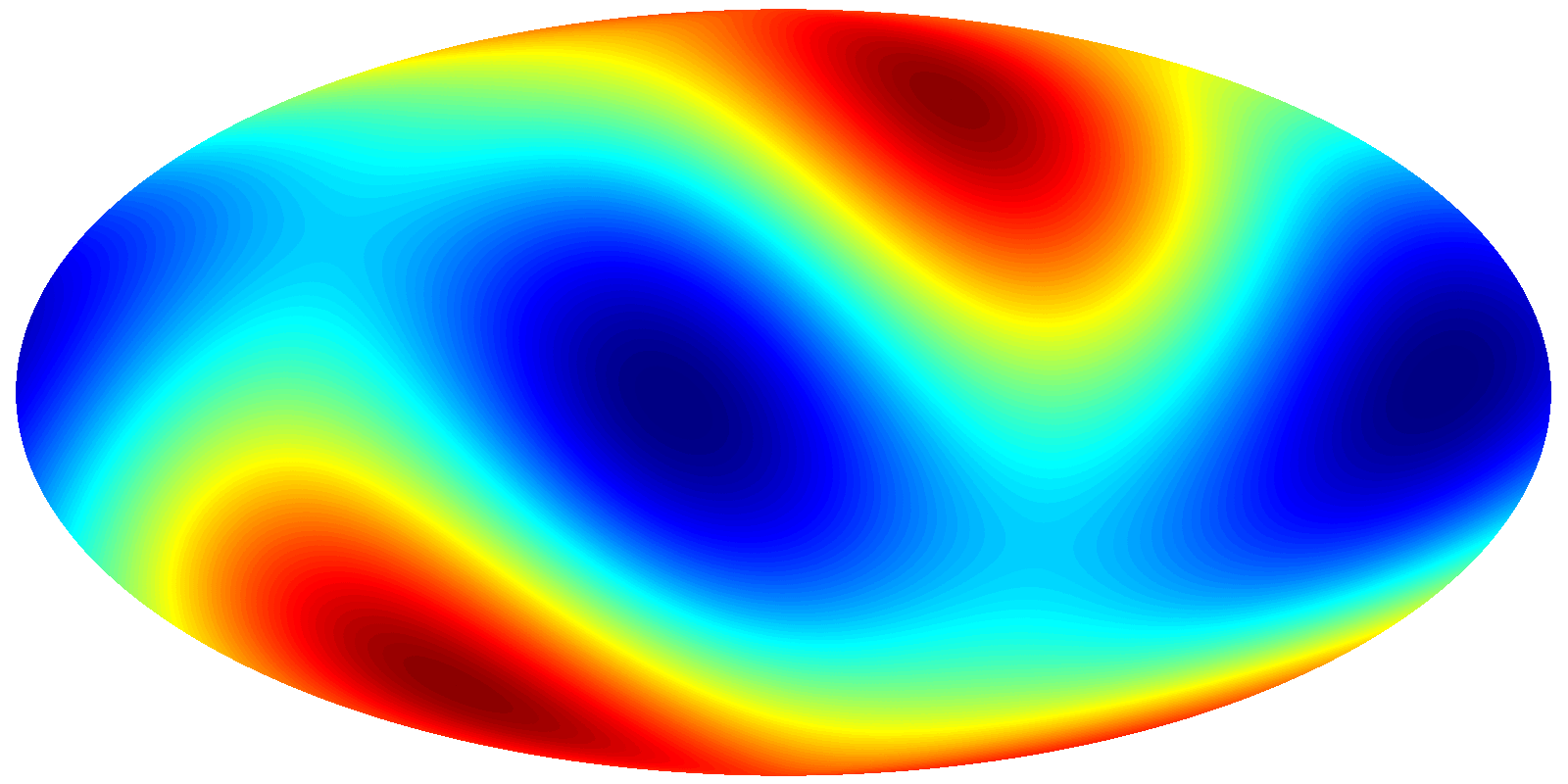}
 \end{minipage}
        \caption{Total quadrupolar angular momentum -- defined by Eq. \eqref{eq:ang-momentum} -- for different directions of the spherical harmonic expansion, for the SMICA 2013 map and galactic coordinates.  Red (blue) regions denote the highest (lowest) values. The red-most point correspond to the $\boldsymbol{\hat{n}}_2$ direction. The actual angular momentum values (\emph{i.e.}~the scale of the plot) is not relevant here. \emph{Left:} raw map ($Q_{\rm eff} = 0$). \emph{Right:} map corrected for the DQ ($Q_{\rm eff} = 1.67$). \label{fig:n2-SMICA}}
\end{figure}

\begin{figure}[t!]
\begin{minipage}[c]{\textwidth}
    \includegraphics[width=0.5\textwidth]{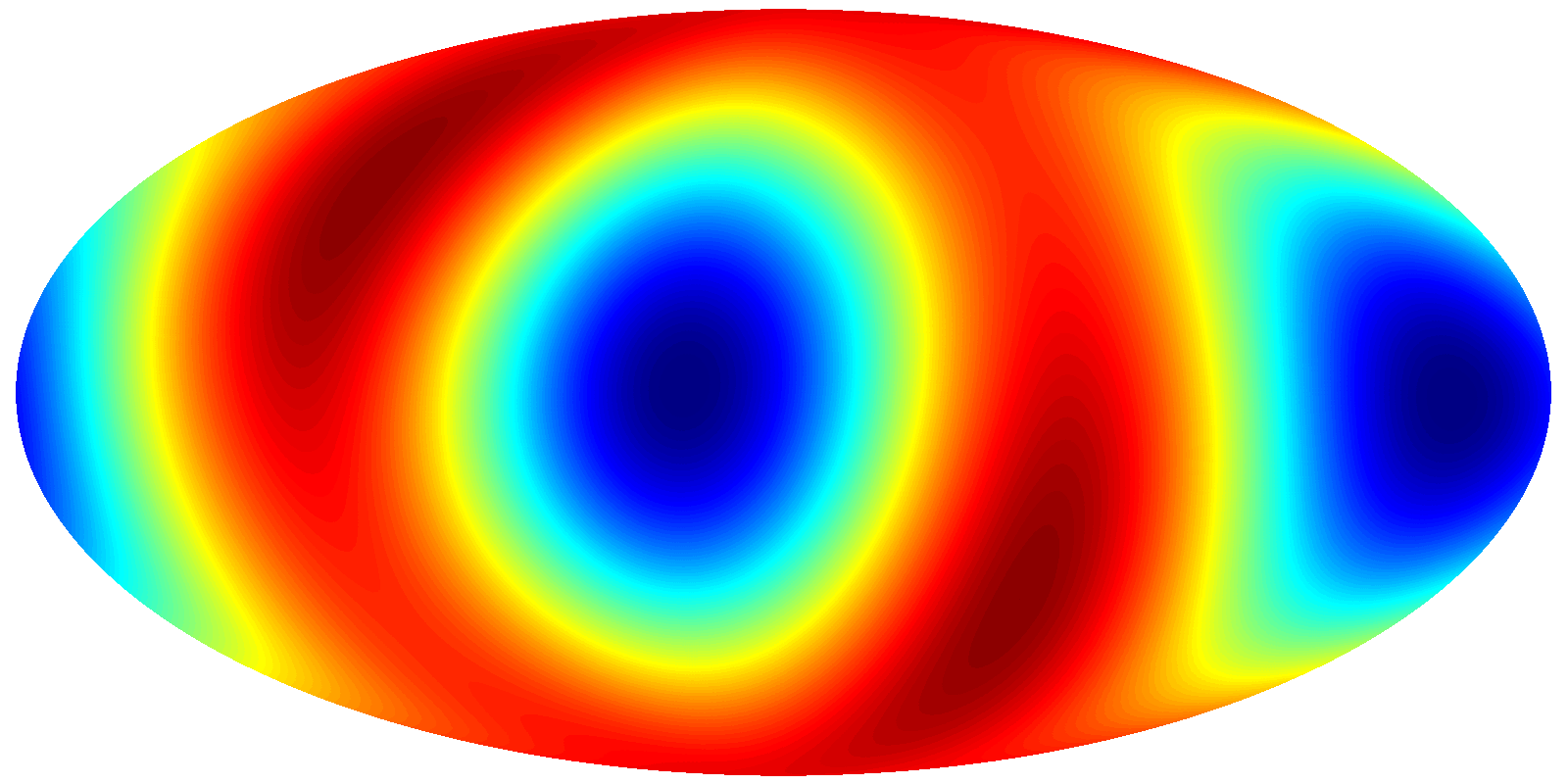}
    \includegraphics[width=0.5\textwidth]{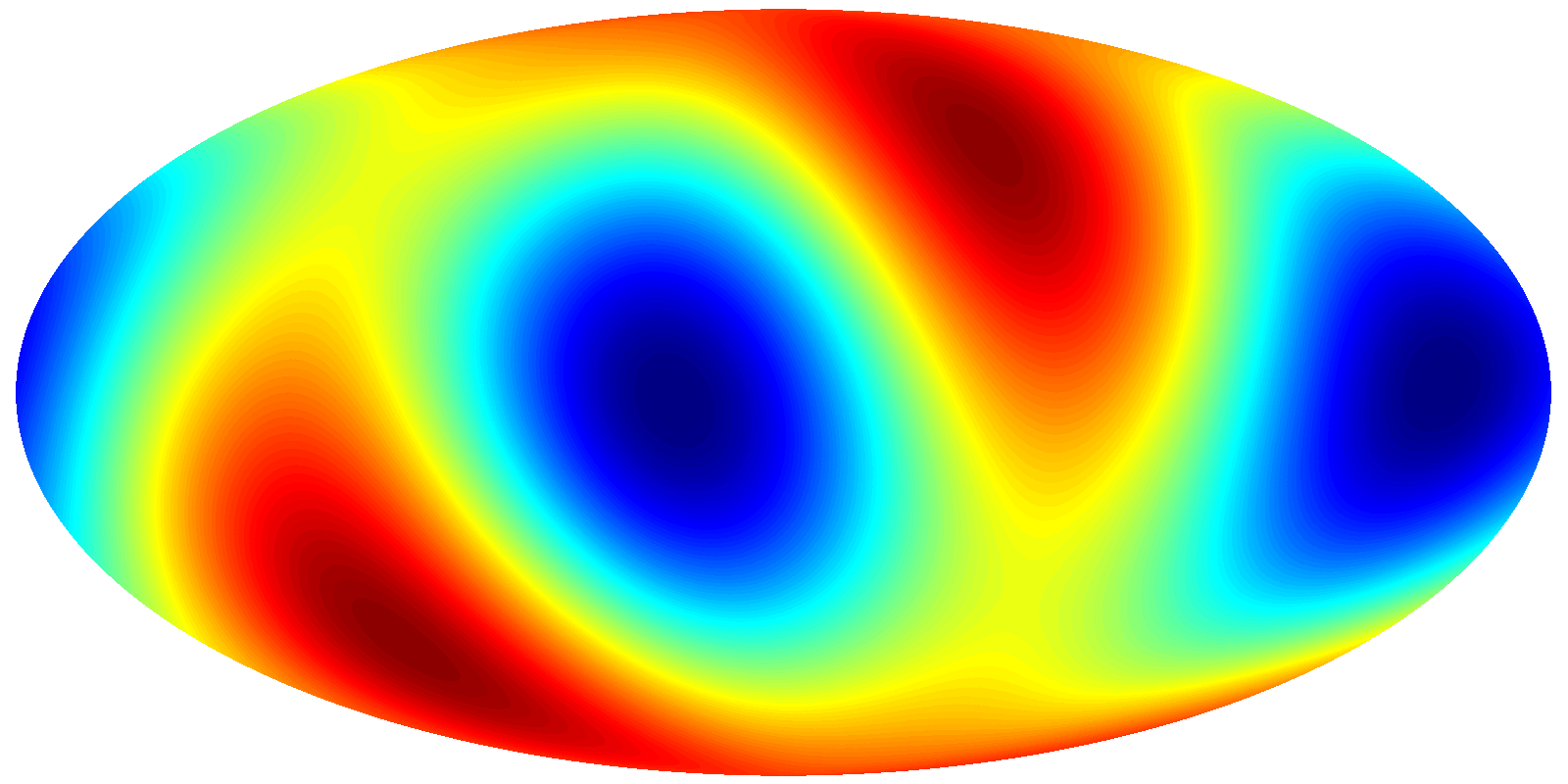}
 \end{minipage}
        \caption{Same as Figure~\ref{fig:n2-SMICA} for the LGMCA-PR1 map (and $Q_{\rm eff} = 2.47$). Note that the difference in $\boldsymbol{\hat{n}}_2$ in this case is much larger, although the overall change in the distribution is not very big.  \label{fig:n2-PR1}}
\end{figure}

In Figures~\ref{fig:n2-SMICA} and~\ref{fig:n2-PR1} we depict the variation of the quadrupolar angular momentum as a function of the direction used to expand the spherical harmonics. The direction that maximizes this value defines $\boldsymbol{\hat{n}}_2$. In both figures, we show on the left hand side the results ignoring the DQ and in the right the properly DQ corrected results. Note that the right-hand plots are in better agreement and that in the PR1 case, which is most affected by the kinetic quadrupole ($Q_{\rm eff} = 2.47$), $\boldsymbol{\hat{n}}_2$ changes substiantially.

We will generally quote our results in $\sigma$ levels (instead of p-values) which are \emph{defined} through the Gaussian distribution:
\begin{equation}\label{eq:pvalues-to-sigma}
    n \sigma \equiv \sqrt{2} \,{\rm Erf}^{-1} \big(1-\textrm{p-value} \big)\,.
\end{equation}
In other words, we use just a straightforward generalization of the commonplace association of $1,\,2,\,3\sigma$ as shorthand notation for $68.3\%,\,95.4\%,\,99.73\%,$ (independently of whether the underlying distribution is really Gaussian) and so forth.

The results for our three alignment statistics are listed in Table~\ref{tab:alignments}. We compare the values in 3 cases: without correcting for the kinetic quadrupole; correcting with the wrong $Q_{\rm eff} = 1$, as done in~\cite{Copi:2013jna,Rassat:2013caa,Rassat:2014yna,Polastri:2015rda}; and finally with the correct result, using the appropriate $Q_{\rm eff}$ for each map. These results are based on a 1--tailed statistical test. In Table~\ref{tab:alignments2} we list the corresponding two-tailed statistics for the case of the appropriate $Q_{\rm eff}$. In fact, if one considers a quadrupole almost exactly orthogonal to the octupole also as an anomaly, then one must use both tails of the distributions for $\boldsymbol{\hat{n}}_2 \cdot \boldsymbol{\hat{n}}_3$. A similar concern applies to $S$ and $T$, whose distributions (taken from interpolation of Figure 6 of~\cite{Copi:2013jna}, realized with $10^6$ inpainted simulations) are peaked around a central value in an almost Gaussian way and so both tails should be considered to be anomalous.\footnote{It is maybe instructive to pose the following sociological question: had the dipole, quadrupole and octupole been observed to be almost exactly perpendicular to each other, would there have been many papers in the literature studying this phenomena? If one thinks this would have happened, the 2--tailed distribution is the correct assessment of the significance as far as the Cosmology community is concerned.}

\begin{table}
\centering
\begin{tabular}{lccccc}
\toprule
\multicolumn{1}{c}{Map} & $Q_{\rm eff}$& $\boldsymbol{\hat{n}}_2 \cdot \boldsymbol{\hat{n}}_3$ (prob)  & S (prob) & T (prob)  \\
\midrule
\multicolumn{5}{c}{Uncorrected values} \\
\midrule
SMICA 2013  & 0 &0.9811 (2.3$\sigma$)   &  0.742 ($2.8\sigma$)& 0.923 ($2.9\sigma$) \\
SMICA 2013 (R14) & 0 & 0.9833 (2.4$\sigma$)  & 0.744 (2.8$\sigma$) & 0.926 ($2.9\sigma$)\\
NILC 2013 & 0 & 0.9901 (2.6$\sigma$)  & 0.734 (2.7$\sigma$)& 0.922 ($2.9\sigma$)\\
NILC 2013 (R14) & 0 & 0.9853 (2.4$\sigma$)  & 0.736 (2.7$\sigma$)& 0.925 ($2.9\sigma$)\\
LGMCA-PR1 (R14) & 0 & 0.1739 (0.2$\sigma$) & 0.558 (1.4$\sigma$) & 0.691 (1.0$\sigma$) \\
LGMCA-WPR1 (R14) & 0 & 0.8851 (1.6$\sigma$)  & 0.696 (2.4$\sigma$) & 0.873 (2.3$\sigma$) \\
\midrule
\multicolumn{5}{c}{Boost-corrected values, incorrect $Q_{\rm eff}$} \\
\midrule
SMICA 2013   & 1 & 0.9964 (2.9$\sigma$)  &  0.775 ($3.1\sigma$) & 0.937 ($3.1\sigma$) \\
SMICA 2013 (R14) & 1 & 0.9973 (3.0$\sigma$) & 0.778 (3.1$\sigma$)& 0.940 ($3.2\sigma$) \\
NILC 2013 & 1 & 0.9987 (3.2$\sigma$) & 0.777 (3.1$\sigma$)& 0.944 ($3.3\sigma$)\\
NILC 2013 (R14) & 1 & 0.9997 (3.6$\sigma$) & 0.772 (3.0$\sigma$)& 0.943 ($3.3\sigma$)\\
LGMCA-PR1 (R14) & 1 & 0.9824 (2.4$\sigma$) & 0.648 (2.0$\sigma$) & 0.838 (2.0$\sigma$) \\
LGMCA-WPR1 (R14) & 1 & 0.9792 (2.3$\sigma$) & 0.747 (2.8$\sigma$)& 0.906 (2.6$\sigma$)\\
\midrule
\multicolumn{5}{c}{Boost-corrected values, correct $Q_{\rm eff}$} \\
\midrule
SMICA 2013 & 1.67 & 0.9990 (3.3$\sigma$)  &  0.788 ($3.2\sigma$) & 0.940 ($3.3\sigma$) \\
SMICA 2013 (R14) & 1.67 & 0.9985 (3.2$\sigma$)  &  0.791 ($3.2\sigma$) & 0.943 ($3.3\sigma$) \\
NILC 2013 & 1.7 (*) & 0.9974 (3.0$\sigma$)  & 0.794 (3.2$\sigma$)& 0.949 ($3.4\sigma$)\\
NILC 2013 (R14) & 1.7 (*) & 0.9990 (3.3$\sigma$)  & 0.786 (3.2$\sigma$)& 0.947 ($3.3\sigma$)\\
LGMCA-PR1 (R14) & 2.47 & 0.9844 (2.4$\sigma$) & 0.752 (2.9$\sigma$) & 0.927 (2.9$\sigma$) \\
LGMCA-WPR1 (R14) & 2.05 & 0.9975 (3.0$\sigma$)  & 0.806 (3.4$\sigma$)& 0.948 ($3.4\sigma$)\\
\end{tabular}
    \medskip
    \caption{1--tailed statistics for the quadrupole-octupole alignments. For SMICA we use both the public map and the one in~\cite{Rassat:2014yna}, to depict the effect of different inpainting techniques. The use of a single tail of the distributions is the correct approach if one considers a quadrupole almost exactly orthogonal to the octupole (and also to the dipole, in the case of $S$ and $T$) as not anomalous. For $Q_{\rm eff} = 1$ we reproduce the results in~\cite{Copi:2013jna,Polastri:2015rda}. (*) For NILC 2013 we were unable to compute $Q_{\rm eff}$ as the different needlet weights were not made available. We instead rely on the number quoted in~\cite{Ade:2013nlj}.}
    \label{tab:alignments}
\end{table}

Having computed the relevant $Q_{\rm eff}$ also has an impact on the amplitude of the quadrupole and so we also show results for the ``low quadrupole'' anomaly. We compute the significance assuming as usual a $\chi^2$ distribution with 5 degrees of freedom and centered around the theoretical estimate, computed using the best-fit $\Lambda$CDM parameters. For Planck SMICA 2013, this amounts to 1170$\,(\mu K)^2$. We list the results in Table~\ref{tab:low-quad}, both using a 1-tailed and 2-tailed statistics. It is clear that the quadrupole is in itself only at most marginally anomalous;  in the case of allowing also the possibility of having a high quadrupole (which we see no reason not to) it is not anomalous at all. The effect of DQ correction turns out to be an overall increase of around $0.2\sigma$ in all the significance levels in both statistics. This seems to be of the same order as the difference among the different map making techniques, which differ mostly in their treatment of the regions close to the galactic plane.

We now also briefly discuss the possibility of other astrophysical (non-primordial) effects contributing to the quadrupole and octupole. Two such effects were studied in~\cite{Rassat:2014yna}: the ISW contribution (which affects the low-$\ell$ modes of the CMB) and the kinetic Sunyaev-Zel'dovich effect (kSZ) from CMB photons scattering off moving free electrons in the local group. They found that the former made a significant impact on these large scale-modes, whereas the latter effect was too small. For the low-quadrupole significance the net effect was small because ISW subtraction affects both  the measured value and the theoretical expectations.  For the $\ell = 2,\,3$ alignments, however, they were claimed to become insignificant after ISW corrections. This would imply that the alignments, if not a statistical fluke, might be fundamentally due to some interactions of the photons with large-scale structure instead of due to some primordial physics. Nevertheless the amplitude of the ISW effect was estimated in~\cite{Ade:2015dva} to be significantly smaller and in agreement with the theoretical estimate of~\cite{2010A&A...520A.101H}, which if confirmed would make its effect on the alignments smaller than the estimate of~\cite{Rassat:2014yna}.

\begin{table}
\centering
\begin{tabular}{lcccc}
\toprule
\multicolumn{1}{c}{Map} & $\boldsymbol{\hat{n}}_2 \cdot \boldsymbol{\hat{n}}_3$ (prob)  & S (prob) & T (prob)  \\
\midrule
\multicolumn{4}{c}{Boost-corrected values, correct $Q_{\rm eff}$} \\
\midrule
SMICA 2013  & 0.9990 (3.1$\sigma$) &  0.788 ($3.0\sigma$) & 0.940 ($2.6\sigma$) \\
SMICA 2013 (R14)  & 0.9985 (3.0$\sigma$) &  0.791 ($3.0\sigma$) & 0.943 ($2.7\sigma$) \\
NILC 2013  & 0.9974 (2.8$\sigma$)  & 0.794 (3.1$\sigma$)& 0.949 ($2.8\sigma$)\\
NILC 2013 (R14) & 0.9990  (3.1$\sigma$) & 0.786 (3.0$\sigma$)& 0.947 ($2.8\sigma$)\\
LGMCA-PR1 (R14) & 0.9844 (2.2$\sigma$) & 0.752 (2.6$\sigma$) & 0.927 (2.4$\sigma$) \\
LGMCA-WPR1 (R14) & 0.9975 (2.8$\sigma$) & 0.806 (3.2$\sigma$)& 0.948 ($2.8\sigma$)\\
\end{tabular}
    \medskip
    \caption{2--tailed statistics for the quadrupole-octupole alignments, using the appropriate $Q(\nu)$. The use of both tails of the distributions is the correct approach if one also considers a quadrupole almost exactly orthogonal to the octupole. In the case of $S$ and $T$ the distribution is bell shaped and so both small and large values are to be considered anomalous.}
    \label{tab:alignments2}
\end{table}

\begin{table}
\centering
\begin{tabular}{lcccc}
\toprule
\multicolumn{1}{c}{Map} & raw $C_2\;(\mu K)^2$ & corrected $C_2\;(\mu K)^2$ & 1--tail prob & 2--tail prob  \\
\midrule
SMICA 2013  & 251 & 228 & $2.1\sigma$ & $1.1\sigma$ \\
SMICA 2013 (R14) & 235 & 214 & $2.2\sigma$ & $1.2\sigma$ \\
NILC 2013 & 219  & 196 & 2.2$\sigma$& $1.2\sigma$\\
NILC 2013 (R14) & 230  & 209 & 2.2$\sigma$& $1.2\sigma$\\
LGMCA-PR1 (R14) &284 & 224 & 2.1$\sigma$ & $1.1\sigma$ \\
LGMCA-WPR1 (R14) &189 & 158 &  2.4$\sigma$& $1.4\sigma$\\
\end{tabular}
    \medskip
    \caption{1 and 2--tailed statistics for the low-quadrupole, using the appropriate $Q(\nu)$. The use of both tails of the distributions is the correct approach if one also considers a high-quadrupole as an anomaly. The significance of the low-quadrupole in itself is marginally significant only if one neglects this.}
    \label{tab:low-quad}
\end{table}

\section{Conclusions}

We have analyzed the impact of the Doppler effect on the quadrupole for WMAP and Planck and we have found that it induces a correction factor which is frequency dependent and large at high frequencies. While for WMAP, which operated at smaller frequencies, such a factor is of small importance, for Planck this is not the case.

The effective correction factor for Planck is equal to: 1.67 for the SMICA 2013 map, 1.7 for NILC 2013, and 2.47 and 2.05 for the PR1 and WPR1 full-sky reconstructions with the LGMCA component separation method of~\cite{Rassat:2014yna}. We have applied such a factor to the analyses of~\cite{Copi:2013jna,Rassat:2014yna}  finding a variable change in the significance of,  in some cases as large as $1\sigma$.

After properly correcting for the DQ we find that the quadrupole-octupole alignment in the released inpainted versions of SMICA 2013 and NILC 2013 maps (which replaces $3\%$ of the sky) as well as the full-sky WPR1 map are significant enough to be taken in serious consideration as a possible challenge to $\Lambda$CDM. These results are also found in alternative inpaintings of SMICA and NILC, the latter of which masks a larger $8\%$ portion of the sky. On the other hand the full sky PR1 map  and the values quoted in~\cite{Ade:2013nlj}  (based on different versions of SMICA and NILC, which are not public, treated with an inpainting performed on much larger region of the sky, to wit $27\%$) yield lower values, even after the correct subtraction of the DQ. They lead to $2.2\sigma$, $2.4\sigma$ and $2.3\sigma$ (2-tailed statistics), respectively (where we made direct use of the values quoted in~\cite{Ade:2013nlj}). This emphasizes the need for a better understanding of the assumptions and procedures of how the galactic region is treated, \emph {i.e.} by masking and inpainting with various assumptions or by direct reconstruction.

The frequency dependency of the quadrupole has been so far mostly neglected in the literature. We have shown in fact in~\cite{Quartin:2015kaa} that it was also ignored in the calibration procedure of Planck 2015 data (which is based on the orbital dipole) even though it is of the same order of magnitude as other effects considered. An inclusion of the $Q(\nu)$ term could thus improve the calibration, for instance suppressing the leakage of temperature into polarization modes.

It is important to realize that the CMB largest scales are completely cosmic-variance dominated and so, decreasing the experimental noise yields no improvements. These can come only through better understanding of the systematics, reduction of the masked portion of the sky or measurements of large-scale polarization. In effect, the DQ is also relevant for polarization, so its proper treatment will also be important there.

It is also relevant  to stress that even a small change in the weights of the different channels would result in a potentially large change in the factor $Q_{\rm eff}$ and therefore in the direction of the quadrupole. So, in the presence of significant uncertainties in foreground subtraction, which would change the relative weights, one has to be very careful in applying the correct $Q_{\rm eff}$ factor to the quadrupole.

\acknowledgments
We thank Alessandro Gruppuso, Carlos Hern\'andez-Monteagudo, Paolo Natoli and Dominik Schwarz for useful discussions and comments.

\bibliographystyle{JHEP2015}
\bibliography{asymmetry}

\providecommand{\href}[2]{#2}\begingroup\raggedright\begin{thebibliography}{10}

\bibitem{Bennett:2010jb}
C.~Bennett, R.~Hill, G.~Hinshaw, D.~Larson, K.~Smith, et~al., {\it {Seven-Year
  Wilkinson Microwave Anisotropy Probe (WMAP) Observations: Are There Cosmic
  Microwave Background Anomalies?}},  {\em Astrophys.J.Suppl.} {\bf 192} (2011)
  17, [\href{http://xxx.lanl.gov/abs/1001.4758}{{\tt arXiv:1001.4758}}].

\bibitem{Ade:2013nlj}
{\bf Planck} Collaboration, P.~Ade et~al., {\it {Planck 2013 results. XXIII.
  Isotropy and statistics of the CMB, v3}},
  \href{http://xxx.lanl.gov/abs/1303.5083}{{\tt arXiv:1303.5083}}.

\bibitem{Challinor:2002zh}
A.~Challinor and F.~van Leeuwen, {\it {Peculiar velocity effects in high
  resolution microwave background experiments}},  {\em Phys.Rev.} {\bf D65}
  (2002) 103001, [\href{http://xxx.lanl.gov/abs/astro-ph/0112457}{{\tt
  astro-ph/0112457}}].

\bibitem{Burles:2006xf}
S.~Burles and S.~Rappaport, {\it {Aberration of the Cosmic Microwave
  Background}},  {\em Astrophys.J.} {\bf 641} (2006) L1--L4,
  [\href{http://xxx.lanl.gov/abs/astro-ph/0601559}{{\tt astro-ph/0601559}}].

\bibitem{Notari:2011sb}
A.~Notari and M.~Quartin, {\it {Measuring our Peculiar Velocity by
  'Pre-deboosting' the CMB}},  {\em JCAP} {\bf 1202} (2012) 026,
  [\href{http://xxx.lanl.gov/abs/1112.1400}{{\tt arXiv:1112.1400}}].

\bibitem{Kosowsky:2010jm}
A.~Kosowsky and T.~Kahniashvili, {\it {The Signature of Proper Motion in the
  Microwave Sky}},  {\em Phys.Rev.Lett.} {\bf 106} (2011) 191301,
  [\href{http://xxx.lanl.gov/abs/1007.4539}{{\tt arXiv:1007.4539}}].

\bibitem{Amendola:2010ty}
L.~Amendola, R.~Catena, I.~Masina, A.~Notari, M.~Quartin, et~al., {\it
  {Measuring our peculiar velocity on the CMB with high-multipole off-diagonal
  correlations}},  {\em JCAP} {\bf 1107} (2011) 027,
  [\href{http://xxx.lanl.gov/abs/1008.1183}{{\tt arXiv:1008.1183}}].

\bibitem{Aghanim:2013suk}
{\bf Planck} Collaboration, N.~Aghanim et~al., {\it {Planck 2013 results.
  XXVII. Doppler boosting of the CMB: Eppur si muove}},
  \href{http://xxx.lanl.gov/abs/1303.5087}{{\tt arXiv:1303.5087}}.

\bibitem{Bennett:1996ce}
C.~Bennett, A.~Banday, K.~Gorski, G.~Hinshaw, P.~Jackson, et~al., {\it {Four
  year COBE DMR cosmic microwave background observations: Maps and basic
  results}},  {\em Astrophys.J.} {\bf 464} (1996) L1--L4,
  [\href{http://xxx.lanl.gov/abs/astro-ph/9601067}{{\tt astro-ph/9601067}}].

\bibitem{Hinshaw:2003ex}
{\bf WMAP} Collaboration, G.~Hinshaw et~al., {\it {First year Wilkinson
  Microwave Anisotropy Probe (WMAP) observations: The Angular power spectrum}},
   {\em Astrophys.J.Suppl.} {\bf 148} (2003) 135,
  [\href{http://xxx.lanl.gov/abs/astro-ph/0302217}{{\tt astro-ph/0302217}}].

\bibitem{Spergel:2003cb}
{\bf WMAP} Collaboration, D.~Spergel et~al., {\it {First year Wilkinson
  Microwave Anisotropy Probe (WMAP) observations: Determination of cosmological
  parameters}},  {\em Astrophys.J.Suppl.} {\bf 148} (2003) 175--194,
  [\href{http://xxx.lanl.gov/abs/astro-ph/0302209}{{\tt astro-ph/0302209}}].

\bibitem{Tegmark:2003ve}
M.~Tegmark, A.~de~Oliveira-Costa, and A.~Hamilton, {\it {A high resolution
  foreground cleaned CMB map from WMAP}},  {\em Phys.Rev.} {\bf D68} (2003)
  123523, [\href{http://xxx.lanl.gov/abs/astro-ph/0302496}{{\tt
  astro-ph/0302496}}].

\bibitem{Efstathiou:2003wr}
G.~Efstathiou, {\it {The Statistical significance of the low CMB multipoles}},
  {\em Mon.Not.Roy.Astron.Soc.} {\bf 346} (2003) L26,
  [\href{http://xxx.lanl.gov/abs/astro-ph/0306431}{{\tt astro-ph/0306431}}].

\bibitem{Efstathiou:2003tv}
G.~Efstathiou, {\it {A Maximum likelihood analysis of the low CMB multipoles
  from WMAP}},  {\em Mon.Not.Roy.Astron.Soc.} {\bf 348} (2004) 885,
  [\href{http://xxx.lanl.gov/abs/astro-ph/0310207}{{\tt astro-ph/0310207}}].

\bibitem{Ade:2013kta}
{\bf Planck} Collaboration, P.~Ade et~al., {\it {Planck 2013 results. XV. CMB
  power spectra and likelihood}},  {\em Astron.Astrophys.} {\bf 571} (2014)
  A15, [\href{http://xxx.lanl.gov/abs/1303.5075}{{\tt arXiv:1303.5075}}].

\bibitem{Hinshaw:1996uq}
G.~Hinshaw, A.~Banday, C.~Bennett, K.~Gorski, A.~Kogut, et~al., {\it {Band
  power spectra in the COBE DMR 4-year anisotropy maps}},  {\em Astrophys.J.}
  {\bf 464} (1996) L17--L20,
  [\href{http://xxx.lanl.gov/abs/astro-ph/9601058}{{\tt astro-ph/9601058}}].

\bibitem{Bond:1998zw}
J.~Bond, A.~H. Jaffe, and L.~Knox, {\it {Estimating the power spectrum of the
  cosmic microwave background}},  {\em Phys.Rev.} {\bf D57} (1998) 2117--2137,
  [\href{http://xxx.lanl.gov/abs/astro-ph/9708203}{{\tt astro-ph/9708203}}].

\bibitem{Ade:2013zuv}
{\bf Planck} Collaboration, P.~Ade et~al., {\it {Planck 2013 results. XVI.
  Cosmological parameters}},  {\em Astron.Astrophys.} {\bf 571} (2014) A16,
  [\href{http://xxx.lanl.gov/abs/1303.5076}{{\tt arXiv:1303.5076}}].

\bibitem{Kamionkowski:2002nd}
M.~Kamionkowski and L.~Knox, {\it {Aspects of the cosmic microwave background
  dipole}},  {\em Phys.Rev.} {\bf D67} (2003) 063001,
  [\href{http://xxx.lanl.gov/abs/astro-ph/0210165}{{\tt astro-ph/0210165}}].

\bibitem{Chluba:2004cn}
J.~Chluba and R.~Sunyaev, {\it {Superposition of blackbodies and the dipole
  anisotropy: A Possibility to calibrate CMB experiments}},  {\em
  Astron.Astrophys.} {\bf 424} (2003) 389--408,
  [\href{http://xxx.lanl.gov/abs/astro-ph/0404067}{{\tt astro-ph/0404067}}].

\bibitem{Sunyaev:2013coa}
R.~A. Sunyaev and R.~Khatri, {\it {Motion induced second order temperature and
  y-type anisotropies after the subtraction of linear dipole in the CMB maps}},
   {\em JCAP} {\bf 1303} (2013) 012,
  [\href{http://xxx.lanl.gov/abs/1302.6571}{{\tt arXiv:1302.6571}}].

\bibitem{Eriksen:2003db}
H.~Eriksen, F.~Hansen, A.~Banday, K.~Gorski, and P.~Lilje, {\it {Asymmetries in
  the Cosmic Microwave Background anisotropy field}},  {\em Astrophys.J.} {\bf
  605} (2004) 14--20, [\href{http://xxx.lanl.gov/abs/astro-ph/0307507}{{\tt
  astro-ph/0307507}}].

\bibitem{Hansen:2004mj}
F.~K. Hansen, P.~Cabella, D.~Marinucci, and N.~Vittorio, {\it {Asymmetries in
  the local curvature of the WMAP data}},  {\em Astrophys.J.} {\bf 607} (2004)
  L67--L70, [\href{http://xxx.lanl.gov/abs/astro-ph/0402396}{{\tt
  astro-ph/0402396}}].

\bibitem{Hansen:2004vq}
F.~K. Hansen, A.~Banday, and K.~Gorski, {\it {Testing the cosmological
  principle of isotropy: Local power spectrum estimates of the WMAP data}},
  {\em Mon.Not.Roy.Astron.Soc.} {\bf 354} (2004) 641--665,
  [\href{http://xxx.lanl.gov/abs/astro-ph/0404206}{{\tt astro-ph/0404206}}].

\bibitem{Eriksen:2007pc}
H.~K. Eriksen, A.~Banday, K.~Gorski, F.~Hansen, and P.~Lilje, {\it
  {Hemispherical power asymmetry in the three-year Wilkinson Microwave
  Anisotropy Probe sky maps}},  {\em Astrophys.J.} {\bf 660} (2007) L81--L84,
  [\href{http://xxx.lanl.gov/abs/astro-ph/0701089}{{\tt astro-ph/0701089}}].

\bibitem{Hoftuft:2009rq}
J.~Hoftuft, H.~Eriksen, A.~Banday, K.~Gorski, F.~Hansen, et~al., {\it
  {Increasing evidence for hemispherical power asymmetry in the five-year WMAP
  data}},  {\em Astrophys.J.} {\bf 699} (2009) 985--989,
  [\href{http://xxx.lanl.gov/abs/0903.1229}{{\tt arXiv:0903.1229}}].

\bibitem{Notari:2013iva}
A.~Notari, M.~Quartin, and R.~Catena, {\it {CMB Aberration and Doppler Effects
  as a Source of Hemispherical Asymmetries}},  {\em JCAP} {\bf 1403} (2014)
  019, [\href{http://xxx.lanl.gov/abs/1304.3506}{{\tt arXiv:1304.3506}}].

\bibitem{Quartin:2014yaa}
M.~Quartin and A.~Notari, {\it {On the significance of power asymmetries in
  Planck CMB data at all scales}},  {\em JCAP} {\bf 1501} (2015), no.~01 008,
  [\href{http://xxx.lanl.gov/abs/1408.5792}{{\tt arXiv:1408.5792}}].

\bibitem{Hansen:2008ym}
F.~Hansen, A.~Banday, K.~Gorski, H.~Eriksen, and P.~Lilje, {\it {Power
  Asymmetry in Cosmic Microwave Background Fluctuations from Full Sky to
  Sub-degree Scales: Is the Universe Isotropic?}},  {\em Astrophys.J.} {\bf
  704} (2009) 1448--1458, [\href{http://xxx.lanl.gov/abs/0812.3795}{{\tt
  arXiv:0812.3795}}].

\bibitem{Bielewicz:2004en}
P.~Bielewicz, K.~Gorski, and A.~Banday, {\it {Low order multipole maps of CMB
  anisotropy derived from WMAP}},  {\em Mon.Not.Roy.Astron.Soc.} {\bf 355}
  (2004) 1283, [\href{http://xxx.lanl.gov/abs/astro-ph/0405007}{{\tt
  astro-ph/0405007}}].

\bibitem{deOliveiraCosta:2003pu}
A.~de~Oliveira-Costa, M.~Tegmark, M.~Zaldarriaga, and A.~Hamilton, {\it {The
  Significance of the largest scale CMB fluctuations in WMAP}},  {\em
  Phys.Rev.} {\bf D69} (2004) 063516,
  [\href{http://xxx.lanl.gov/abs/astro-ph/0307282}{{\tt astro-ph/0307282}}].

\bibitem{Copi:2003kt}
C.~J. Copi, D.~Huterer, and G.~D. Starkman, {\it {Multipole vectors - A New
  representation of the CMB sky and evidence for statistical anisotropy or
  non-Gaussianity at $2 \le l \le 8$}},  {\em Phys.Rev.} {\bf D70} (2004)
  043515, [\href{http://xxx.lanl.gov/abs/astro-ph/0310511}{{\tt
  astro-ph/0310511}}].

\bibitem{Copi:2013jna}
C.~J. Copi, D.~Huterer, D.~J. Schwarz, and G.~D. Starkman, {\it {Large-scale
  alignments from WMAP and Planck}},
  \href{http://xxx.lanl.gov/abs/1311.4562}{{\tt arXiv:1311.4562}}.

\bibitem{Rassat:2013caa}
A.~Rassat, J.~L. Starck, and F.~X. Dupe, {\it {Removal of two large scale
  Cosmic Microwave Background anomalies after subtraction of the Integrated
  Sachs Wolfe effect}},  \href{http://xxx.lanl.gov/abs/1303.4727}{{\tt
  arXiv:1303.4727}}.

\bibitem{Rassat:2014yna}
A.~Rassat, J.-L. Starck, P.~Paykari, F.~Sureau, and J.~Bobin, {\it {Planck CMB
  Anomalies: Astrophysical and Cosmological Secondary Effects and the Curse of
  Masking}},  {\em JCAP} {\bf 1408} (2014) 006,
  [\href{http://xxx.lanl.gov/abs/1405.1844}{{\tt arXiv:1405.1844}}].

\bibitem{Polastri:2015rda}
L.~Polastri, A.~Gruppuso, and P.~Natoli, {\it {CMB low multipole alignments in
  the $\mathbf{\Lambda}$CDM and Dipolar models}},  {\em JCAP} {\bf 1504}
  (2015), no.~04 018, [\href{http://xxx.lanl.gov/abs/1503.01611}{{\tt
  arXiv:1503.01611}}].

\bibitem{Peebles:1968}
P.~J. {Peebles} and D.~T. {Wilkinson}, {\it {Comment on the Anisotropy of the
  Primeval Fireball}},  {\em Physical Review} {\bf 174} (Oct., 1968)
  2168--2168.

\bibitem{Ade:2013sjv}
{\bf Planck} Collaboration, P.~Ade et~al., {\it {Planck 2013 results. I.
  Overview of products and scientific results}},  {\em Astron.Astrophys.} {\bf
  571} (2014) A1, [\href{http://xxx.lanl.gov/abs/1303.5062}{{\tt
  arXiv:1303.5062}}].

\bibitem{Ade:2013hta}
{\bf Planck} Collaboration, P.~Ade et~al., {\it {Planck 2013 results. XII.
  Diffuse component separation}},  {\em Astron.Astrophys.} {\bf 571} (2014)
  A12, [\href{http://xxx.lanl.gov/abs/1303.5072}{{\tt arXiv:1303.5072}}].

\bibitem{Adam:2015wua}
{\bf Planck} Collaboration, R.~Adam et~al., {\it {Planck 2015 results. X.
  Diffuse component separation: Foreground maps}},
  \href{http://xxx.lanl.gov/abs/1502.01588}{{\tt arXiv:1502.01588}}.

\bibitem{Adam:2015tpy}
{\bf Planck} Collaboration, R.~Adam et~al., {\it {Planck 2015 results. IX.
  Diffuse component separation: CMB maps}},
  \href{http://xxx.lanl.gov/abs/1502.05956}{{\tt arXiv:1502.05956}}.

\bibitem{Bobin:2014mja}
J.~Bobin, F.~Sureau, J.~L. Starck, A.~Rassat, and P.~Paykari, {\it {Joint
  Planck and WMAP CMB Map Reconstruction}},  {\em Astron.Astrophys.} {\bf 563}
  (2014) A105, [\href{http://xxx.lanl.gov/abs/1401.6016}{{\tt
  arXiv:1401.6016}}].

\bibitem{BenoitLevy:2013bc}
A.~Benoit-Levy, T.~Dechelette, K.~Benabed, J.-F. Cardoso, D.~Hanson, et~al.,
  {\it {Full-sky CMB lensing reconstruction in presence of sky-cuts}},  {\em
  Astron.Astrophys.} {\bf 555} (2013) A37,
  [\href{http://xxx.lanl.gov/abs/1301.4145}{{\tt arXiv:1301.4145}}].

\bibitem{Ade:2013ktc}
{\bf Planck} Collaboration, P.~Ade et~al., {\it {Planck 2013 results. I.
  Overview of products and scientific results}},
  \href{http://xxx.lanl.gov/abs/1303.5062}{{\tt arXiv:1303.5062}}.

\bibitem{2014A&A...571A...5P}
{Planck Collaboration}, N.~{Aghanim}, C.~{Armitage-Caplan}, M.~{Arnaud},
  M.~{Ashdown}, F.~{Atrio-Barandela}, J.~{Aumont}, C.~{Baccigalupi}, A.~J.
  {Banday}, R.~B. {Barreiro}, and et~al., {\it {Planck 2013 results. V. LFI
  calibration}},  {\em Astron.Astrophys.} {\bf 571} (Nov., 2014) A5,
  [\href{http://xxx.lanl.gov/abs/1303.5066}{{\tt arXiv:1303.5066}}].

\bibitem{Ade:2013eta}
{\bf Planck} Collaboration, P.~Ade et~al., {\it {Planck 2013 results. VIII. HFI
  photometric calibration and mapmaking}},  {\em Astron.Astrophys.} {\bf 571}
  (2014) A8, [\href{http://xxx.lanl.gov/abs/1303.5069}{{\tt arXiv:1303.5069}}].

\bibitem{Gorski:2004by}
K.~Gorski, E.~Hivon, A.~Banday, B.~Wandelt, F.~Hansen, et~al., {\it {HEALPix -
  A Framework for high resolution discretization, and fast analysis of data
  distributed on the sphere}},  {\em Astrophys.J.} {\bf 622} (2005) 759--771,
  [\href{http://xxx.lanl.gov/abs/astro-ph/0409513}{{\tt astro-ph/0409513}}].

\bibitem{Larson:2014roa}
D.~Larson, J.~Weiland, G.~Hinshaw, and C.~Bennett, {\it {Comparing Planck and
  WMAP: Maps, Spectra, and Parameters}},  {\em Astrophys.J.} {\bf 801} (2015),
  no.~1 9, [\href{http://xxx.lanl.gov/abs/1409.7718}{{\tt arXiv:1409.7718}}].

\bibitem{cosmostat-webpage}
 \url{http://www.cosmostat.org/}.

\bibitem{Ade:2013fta}
{\bf Planck} Collaboration, P.~Ade et~al., {\it {Planck 2013 results. IX. HFI
  spectral response}},  {\em Astron.Astrophys.} {\bf 571} (2014) A9,
  [\href{http://xxx.lanl.gov/abs/1303.5070}{{\tt arXiv:1303.5070}}].

\bibitem{Zonca:2010fx}
A.~Zonca, C.~Franceschet, P.~Battaglia, F.~Villa, A.~Mennella, et~al., {\it
  {Planck-LFI radiometers' spectral response}},  {\em JINST} {\bf 4} (2009)
  T12010, [\href{http://xxx.lanl.gov/abs/1001.4589}{{\tt arXiv:1001.4589}}].

\bibitem{Copi:2005ff}
C.~J. Copi, D.~Huterer, D.~Schwarz, and G.~Starkman, {\it {On the large-angle
  anomalies of the microwave sky}},  {\em Mon.Not.Roy.Astron.Soc.} {\bf 367}
  (2006) 79--102, [\href{http://xxx.lanl.gov/abs/astro-ph/0508047}{{\tt
  astro-ph/0508047}}].

\bibitem{Helling:2006xh}
R.~C. Helling, P.~Schupp, and T.~Tesileanu, {\it {CMB statistical anisotropy,
  multipole vectors and the influence of the dipole}},  {\em Phys.Rev.} {\bf
  D74} (2006) 063004, [\href{http://xxx.lanl.gov/abs/astro-ph/0603594}{{\tt
  astro-ph/0603594}}].

\bibitem{Ade:2015dva}
{\bf Planck} Collaboration, P.~Ade et~al., {\it {Planck 2015 results. XXI. The
  integrated Sachs-Wolfe effect}},
  \href{http://xxx.lanl.gov/abs/1502.01595}{{\tt arXiv:1502.01595}}.

\bibitem{2010A&A...520A.101H}
C.~{Hern{\'a}ndez-Monteagudo}, {\it {Revisiting the WMAP-NVSS angular cross
  correlation. A skeptic's view}},  {\em Astron.Astrophys.} {\bf 520} (Sept.,
  2010) A101, [\href{http://xxx.lanl.gov/abs/0909.4294}{{\tt
  arXiv:0909.4294}}].

\bibitem{Quartin:2015kaa}
M.~Quartin and A.~Notari, {\it {Improving Planck calibration by including
  frequency-dependent relativistic corrections}},
  \href{http://xxx.lanl.gov/abs/1504.04897}{{\tt arXiv:1504.04897}}.

\end{thebibliography}\endgroup

\end{document}